# Practical Power Allocation and Greedy Partner Selection for Cooperative Networks

Hadi Goudarzi
EE School, Sharif University of Tech.
Tehran, Iran
h_goudarzi@ee.sharif.edu

Mohamad Reza Pakravan
EE School, Sharif University of Tech.
Tehran, Iran
pakravan@sharif.edu

*Abstract*- In this paper, we present a novel algorithm for power allocation in the Amplify-and-Forward cooperative communication that minimizes the outage probability with a given value of total power. We present the problem with new formulation and solve the optimal power allocation for a fixed set of partners. The proposed solution provides a direct power allocation scheme with a simple formula that can be also be represented by a simple lookup table which makes it easy for practical implementation. We present simulation results to demonstrate that the performances of the proposed algorithms are very close to results of the previously published iterative optimal power allocation algorithms. We also consider the issue of partner selection in a cooperative network.

*Keywords-component; Cooperative diversity, Amplify and Forward, Outage Probability, Partner Selection.*

## I. INTRODUCTION

Cooperative diversity is a technique that combats the slow fading and shadowing effect in wireless communication channel [1]-[3]. In this technique, the spatially distributed users create an array of antennas to combat slow fading so the achievable rate and capacity of wireless channels will be improved saliently. The technique can also lean to reduction of the required power for transmission.

One of the recent approaches in cooperative diversity problem is minimization of power for constant rate which satisfies a constraint of outage probability or error probability [9]-[13]. In [9], the authors have been used the short term power (minimum power that satisfies capacity constraint of the problem) for solving the problem with constant mean of total transmit power. In [10], the authors focused on the problem of constrained minimization of power but a closed form solution was not presented. Lifetime maximization problem via cooperative nodes in wireless sensor networks is discussed in [11]. In that paper, the minimization of the total power of cooperative nodes has been studied to maximize the network life for a given error probability. In [12], the authors assume the two partners case and solve the minimization of power in the entire network. The adaptive modulation technique is applied in [13] to improve the spectral efficiency of cooperative strategy and minimize the power consumption. In [16], we presented the optimal algorithm for partner selection and power allocation between source and partners to minimize the power in amplify and forward cooperative diversity (AFCD) and we presented these algorithms for Equal Power Allocation (EPA) scheme in [17].

One of the most important problems in cooperative diversity is the strategy of power allocation among users [4]-[6]. Most of the related works focus on the problem to allocate a constant power to the source and its partners to achieve the minimum value of outage probability. The power allocation for the Decode-and-Forward strategy, based on simulation and observation, has been studied in [4]. Also power allocation based on the constrained optimization method has been studied in [5] and [6]. In [5], Annavajjala et al. presented the approximation of the outage probability for the different cooperative diversity in high Signal to Noise Ratio (SNR) regime and then solved the optimal power allocation to minimize the outage probability with a constraint on total transmit power. In [6], the authors presented the optimal power allocation for Amplify and Forward Cooperative Diversity (AFCD) by Channel Side Information (CSI) and without CSI and then presented the opportunistic relaying problem to maximize the gain of the cooperation.

Another important challenge in cooperative diversity is Partner selection and matching. In [7], a partner selection algorithm in an opportunistic relaying form has been proposed. It is assumed that all of the candidates of cooperation are ready to cooperate and in each packet transmission, the best partner will cooperate. The matching and partner selection may be cooperative or greedy. Cooperative matching is performed to minimize the total transmit power of the network nodes (Mahinthan et al. in [12]) or to minimize the maximum transmit power of the nodes (or maximize life time of the network in [11]). In greedy matching, the node does not try to maximize the network performance and tries to maximize its own performance (like the nodes in most Ad-Hoc networks). In [18], the author presents the concept of greedy matching in cooperative networks and compares the benefits of this matching strategy with the other matching strategies.

The optimal power allocation algorithms for AFCD that have been presented in the previous works are complex and need several iterations to reach to the solution. In this paper, we present a new formulation of the problem and present a novel algorithm for power allocation which does not require any iteration. Based on this power allocation algorithm, we present a method with lookup table which with a little memory usage decreases the complexity of this algorithm to calculation of a few simple calculations for the power allocation. By simulation results, we demonstrate that the performances of our proposed algorithms are close to the performance of the optimal power allocation with iteration.

We also consider the problem of partner selection in AFCD for case of only one partner for each node. We present a method for ranking the candidate nodes for cooperation based on its produced performance for cooperation which is related to location of the candidate nodes.

In section II, we express the model of wireless channel and the cooperative strategy which is employed in this paper. We present the power allocation algorithm without iteration in section III. The implementation of the power allocation algorithm with lookup table is presented in section IV. The results of simulations are expressed in section V. In section VI, partner selection method in one partner case is presented and section VII provides the paper conclusion.

## II. SYSTEM MODEL

In this paper, we assume a slow, flat fading wireless channel. In other words, the bandwidth of signal is smaller than coherence bandwidth of channel and the inverse of the rate of transmission is smaller than coherence time of channel. Noting this assumption, the fading coefficient of channel can be assumed unchanged in a few transmission periods. The large scale behavior of channel path loss is modeled with $D^{-\alpha}$ where D is the distance between transmitter and receiver and $\alpha$ is a positive constant between 2 and 6.

Our cooperative diversity strategy is Amplify and Forward (AF) with orthogonal transmission. In this strategy, each node has a few partners and the partners relay the received signals from the source to the destination. Each relay can be a source in other transmission time intervals. In the power allocation algorithm of this paper, we assume that the partners set for transmission is given and in the greedy matching strategy, we assume that the nodes can select only one user from the candidate partners set.

The power allocation strategy of the proposed algorithm is based on the information of the means of the channel coefficients, between source and partners and between partners and destination. It is assumed that the receiver has the information of the instantaneous CSI of the channels and uses the maximum ratio combining (MRC) to detect the source information from the signals of source and partners, but the source does not have full channel information.

## III. POWER ALLOCATION ALGORITHM

### A. Outage Behavior Of AFCD

To explain the behavior of the outage probability in AF strategy, we first have to explain the information term. According to [6], the source destination channel capacity in bits per time slot in AF is given by (1).

$$I = \frac{1}{m+1}\log\left(B_0 + \sum_{i=1}^{m}\frac{A_i B_i}{A_i+B_i+1}\right) \quad (1)$$

Where $B_0$ and $B_i$ denote the SNR of the link between source and destination and SNR of the link between i[th] partner and destination and $A_i$ denotes the SNR of the link between source and i[th] partner. Each of $A_i$, $B_i$ and $B_0$ random variables have an exponential distribution because the amplitude of the channel coefficient has a Rayleigh distribution.

To explain the outage probability behavior of (1), we can approximate each m+1 terms of information by exponential distribution and approximate the outage probability by first term of Taylor series expansion where is obtained by moment generating function technique. By this manner, the approximation of outage probability has the form of (2).

$$P_{out} = Prob\{I<R\} \cong \frac{\left((2^{(m+1)R}-1)N_0\right)^{m+1}}{(m+1)!} \frac{d_{sd}^{\alpha}}{P_s} \prod_{i=1}^{m} \frac{\frac{P_s}{d_{sr_i}^{\alpha}}+\frac{P_{r_i}}{d_{r_id}^{\alpha}}}{\frac{P_s}{d_{sr_i}^{\alpha}} \frac{P_{r_i}}{d_{r_id}^{\alpha}}} \quad (2)$$

Where $P_s$ and $P_{r_i}$ denote the transmit powers of the source and i[th] partner and $d_{sr_i}$ and $d_{r_id}$ denote the distance between the source and i[th] partner and between i[th] partner and destination. This approximation has high accuracy in high SNR because we remove 1 from deficit terms in approximation. This approximation is equal to outage approximation of [5] and [6] which are used for optimal power allocation.

### B. Previous Optimal Power Allocation

Optimal power allocation between a given set of partners and a source is an optimization problem. In this problem, we try to minimize the outage probability of transmission of source to destination with cooperation of partners with a constraint on total transmit power. This problem can be formulated as (3).

$$\min \quad P_{out}\left(P_s + \sum_{i=1}^{m} P_{r_i}\right) \quad (3)$$

$$\text{subject to} \quad P_s + \sum_{i=1}^{m} P_{r_i} \leq P_T \quad (3\text{-}1)$$

$$P_{max} \geq P_s, P_{r_i} \geq 0 \quad (3\text{-}2)$$

This problem had been solved by Karush-Kuhn-Tucker (KKT) method in [5] and [6]. In [6], the authors used a pre-determined value for $P_s$ to simplify the optimization problem and obtained a closed form solution. In this solution, the required power for partners can be obtained from (4).

$$P_{r_i} = \frac{-P_s\left(d_{r_id}/d_{sr_i}\right)^{\alpha} + \sqrt{P_s^2\left(d_{r_id}/d_{sr_i}\right)^{2\alpha} + 4P_s\left(d_{r_id}/d_{sr_i}\right)^{\alpha}\lambda}}{2} \quad (4)$$

Where $\lambda$ is the Lagrange multiplier of the KKT method and can be obtained from the constraint on the total transmit power. The proposed solution of [5] for this optimization problem is (4) by a few changes. In [5], the authors claimed that $\lambda$ must be set to $\frac{P_T}{m+1}$. By this change, the algorithm is changed to a suboptimal algorithm and the number of unknown parameters in (4) is decreased to one. However this solution has not a closed form and requires a numerical optimization technique to implement it and find the optimal power allocation between source and partners to minimize the outage probability.

### C. Our Algorithm

To solve these problems, we present the minimization problem with a new form and solve the power allocation problem by the KKT method again.

In this paper, the transmission term of the source and i[th] partner in (2) is shown by $\lambda_0$ and $\lambda_i$ respectively.

$$\lambda_i = \frac{1}{N_0}\frac{\frac{P_s}{d_{sr_i}^{\alpha}} * \frac{P_{r_i}}{d_{r_id}^{\alpha}}}{\frac{P_s}{d_{sr_i}^{\alpha}}+\frac{P_{r_i}}{d_{r_id}^{\alpha}}} = \lambda_0 * \frac{\frac{d_{sd}^{\alpha}}{d_{sr_i}^{\alpha}} * \frac{P_{r_i}}{d_{r_id}^{\alpha}}}{\frac{P_s}{d_{sr_i}^{\alpha}}+\frac{P_{r_i}}{d_{r_id}^{\alpha}}} = \lambda_0 \lambda_i' \quad (5)$$

Where $\lambda_i'$ is the normalized of $\lambda_i$ by $\lambda_0$.

If the i[th] partner produced transmission term is equal to $\lambda_i$, then its transmission power must be equal to (6).

$$P_{r_i} = \left(\frac{\lambda'_i}{1-\lambda'_i D^\alpha_{sr_i}}\right) * \lambda_0 * d^\alpha_{r_id} = \left(\frac{\lambda_i}{1-\lambda'_i D^\alpha_{sr_i}}\right) * d^\alpha_{r_id} \quad (6)$$

Where $D$ denotes the normalized distance by the distance of source to destination.

Now we can present the outage minimization problem of (3) by new parameters in (7). We note that the minimization of the outage probability in (3) is equivalent to maximization of the multiple of all transmission terms in (7).

$$max \quad (\lambda_0 * \prod_{i=1}^{m} \lambda_i) \quad (7)$$

$$s.t. \quad \lambda_0 * d^\alpha_{sd} + \sum_{i=1}^{m}\left(\left(\frac{\lambda'_i}{1-\lambda'_i D^\alpha_{sr_i}}\right) * \lambda_0 * d^\alpha_{r_id}\right) \leq P_T \quad (7\text{-}1)$$

$$P_{max} \geq P_s, P_{r_i} \geq 0 \quad (7\text{-}2)$$

By using the KKT method in this constrained optimization problem (in [14] and [15]), the solution of the minimization of the outage probability leads to satisfying (m+1) equation of (8). Equation (8) presents the required amount of $\lambda'_i$ for the i$^{th}$ partner where the index i varies from 1 to m (number of partners).

$$\begin{pmatrix} \prod_{i=1}^{m}\lambda_i \\ \vdots \\ \frac{\lambda_0}{\lambda_m}\prod_{i=1}^{m}\lambda_i \end{pmatrix} = \theta \begin{pmatrix} d^2_{sd} + \sum_{i=1}^{m}\left(\frac{-(\lambda'_i)^2 * D^\alpha_{sr_i} * d^\alpha_{r_id}}{\left(1-\lambda'_i D^\alpha_{sr_i}\right)^2}\right) \\ \vdots \\ \frac{1}{\left(1-\lambda'_m D^\alpha_{sr_m}\right)^2} d_{r_md}^2 \end{pmatrix} \quad (8)$$

Where $\theta$ denotes the Lagrange multiplier of the KKT method. The solution of (8) reaches to the required amount of $\lambda'_i$ where expressed in (9).

$$\lambda'_i = \frac{D^\alpha_{sr_i} + \frac{1}{2}\zeta_i D^\alpha_{r_id} - \sqrt{\frac{1}{4}\zeta_i^2 D^{2\alpha}_{r_id} + (1+\zeta_i) D^\alpha_{sr_i} D^\alpha_{r_id}}}{D^{2\alpha}_{sr_i} - D^\alpha_{sr_i} D^\alpha_{r_id}} \quad (9)$$

$$\zeta_i = 1 + \sum_{j=1}^{m}\left(\frac{\lambda_j}{\lambda_0} D^\alpha_{sr_j}\right) - \frac{\lambda_i}{\lambda_0} D^\alpha_{sr_i} \quad (10)$$

When all of the $\lambda'_i$ for the partners have been calculated, $\lambda_0$ can be calculated by (11). This equation presents the required amount for $\lambda_0$ for satisfying the total power constraint in (7-1).

$$\lambda_0 = \frac{P_T}{d^\alpha_{sd} + \sum_{i=1}^{m}\left(\left(\frac{\lambda'_i}{1-\lambda'_i D^\alpha_{sr_i}}\right) * d^\alpha_{r_id}\right)} \quad (11)$$

We can find the optimal power for all partners and source from the values where calculated from (9) and (11) equations and apply them to (6) and the equation of $P_s = \lambda_0 * d^\alpha_{sd}$.

Finding the solution of the optimization problem from (9) and (11) requires a few iterations for finding the parameter $\zeta$. In this algorithm, we use (12) for estimation of $\zeta$ and relax the problem from iteration. The equation (12) is obtained from setting $\zeta$ to 1 in the first step and calculating $\zeta$ in the second step. We note that this estimation is general and is not related to a specific condition.

$$\zeta = 1 + \sum_{\substack{j=1 \\ j \neq i}}^{m}\left(\frac{D^\alpha_{sr_j} + \frac{1}{2}D^\alpha_{r_jd} - \sqrt{\frac{1}{4}D^{2\alpha}_{r_jd} + 2D^\alpha_{sr_j} D^\alpha_{r_jd}}}{D^{2\alpha}_{sr_j} - D^\alpha_{sr_j} D^\alpha_{r_jd}}\right) \quad (12)$$

Now, we can present the power allocation algorithm without any iteration in the following steps:

1- Calculate $\zeta$ from (12) and calculate (9) for all partners.
2- Calculate $\lambda_0$ from (11) (and calculate the source power).
3- Calculate $P_{r_i}$ from (6).

This algorithm has a closed form and can be implemented by very low computation ($2(m+1)$ equations for $(m+1)$ power levels). This algorithm does not require any iteration for calculating the transmit powers.

The first step of this algorithm is not dependent to constraint total power and can be implemented by a look up table. In the next section, we will present the implementation of this algorithm with lookup table.

## IV. AN ALGORITHM WITH LOOKUP TABLE

The parameter $\zeta$ is dependent to the partners set. If we degrade the accuracy of the estimation of $\zeta_i$ and set it to 1 for $i$ from 1 to m, we can calculate the parameter $\lambda'_i$ for the partners in different locations independent from the partners set where are shown in figure 1. In this figure, the source and destination are placed on (x,y)=(0,0) and (x,y)=(1,0) respectively. We note that for each partner, $\lambda'_i$ can be a value between 0 and $1/D^\alpha_{sr_i}$ i.e. for the partners where are close to destination, $\lambda'_i$ is limited by 1 and this shows that these partners extremely can produce the transmission term (in (2)) equal to source transmission term.

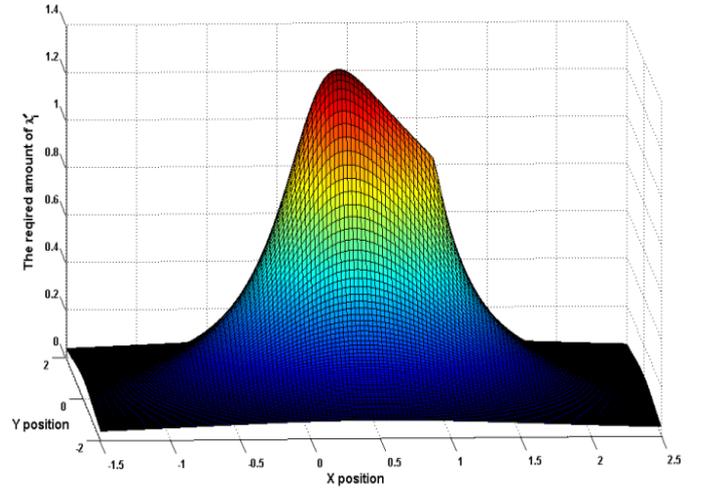

Figure 1. the required $\lambda'_i$ for different position of partners

Figure 1 shows that if the partner is located far from source and destination, the parameter $\lambda'_i$ has a small value. This property shows that we can implement the first step of our presented algorithm by a finite lookup table.

To show the performance of the algorithm with lookup table, we implement this lookup table by a table with 100 values between 0 and 140 which equivalent to 100 bytes. This means that we assume that if the partner is located away from this table we set the zero power for it. By this amount of memory usage, we need to $m+2$ simple calculation (like sum and division) in two other steps of our algorithm.

In the next section we will show that the performance of the presented algorithms (with low computational complexity) is close to Optimal Power Allocation (OPA) with iteration.

## V. SIMULATION RESULTS

### A. Performance of Different Algorithm

In this section, we present the results for comparison between different power allocation algorithms. For this reason, we implement our algorithm (with and without lookup table) with presented steps and OPA with iteration (by Newton method). For comparison of the performance of optimal scheme with Equal Power Allocation (EPA) scheme, we implement two EPA schemes where in the first, the total power is allocated between source and all partners equally and in second, like to [4] for decode and forward, the half of the total power is allocated to source and the remaining half of power is allocated between all partners equally.

Figure 2 and 3 show the outage probability of different power allocation algorithms for different total power. In these simulations, we assume that $d_{sd} = 100m$, $\alpha = 2$, $N_0 = 10^{-4}$ and $R = 1$ bit/s/Hz. In figure 2, two partners located in the normalized distance of $[D_{sr_i}, D_{r_id}] = [1.02, 0.32]$ and $[D_{sr_i}, D_{r_id}] = [0.97, 0.31]$ and In figure 3, one partner located in the normalized distance of $[D_{sr_i}, D_{r_id}] = [0.5, 0.5]$. The probabilities of these figures are calculated with more than 20 million channel realizations (in high SNR) with MATLAB. These figures show that our algorithm results without any iteration are very close to OPA results with iteration and The second presented algorithm with lookup table does not degrade the performance of the system greater than $0.1\ dB$. These figures show that the performance of the second EPA is better than the first EPA in this two relay case. The difference between the outage of the OPA and non-cooperative scheme is growing by increasing the total transmit power. For example for the outage probability of 0.05 the non-cooperative scheme requires $7dB$ more of power than the total power of OPA in one relay case in figure 3.

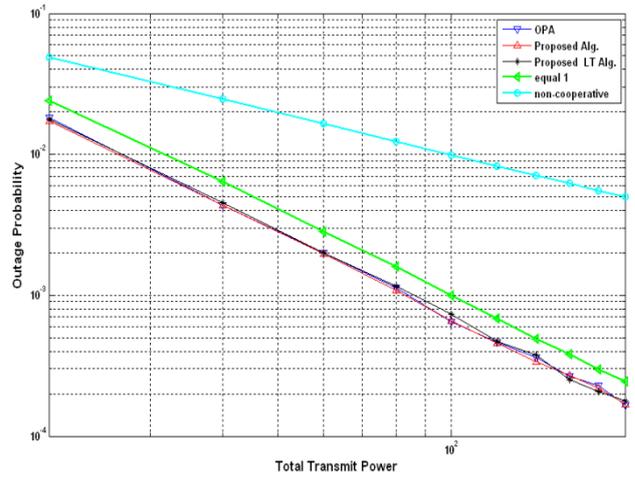

Figure 3- The outage probability of the difference schemes for different total power with one partner

### B. Optimal Number of Partners

The number of cooperative partners is one of the important problems in cooperative diversity with constraint on the total transmit power. Increasing the number of cooperative partners increase the order of diversity also increase the usage of spectrum and decrease the level of the power of the other partners. This means that increasing the number of cooperative partners may does not decrease the outage probability with constraint on the total transmit power. This problem in extreme case changes to the problem of to cooperate or not cooperate (which is discussed in Diversity Multiplexing Tradeoff discussion in [2]).

The problem of finding the optimal number of cooperative partners is related to many parameters like $P_T$, $R$, $d_{sd}$ and the number and locations of the candidate partners. For example, if the number of candidate partners is 2, the number of cooperative partners is not greater than 2 and if one of the candidate partners has a bad channel to destination, our solution of the number of cooperative partners may be changed to the problem of to cooperate with one user or not.

To relax the dependency of the problem to the state of the candidate partners, we assume that the source has infinite candidate partners in one mediocre location ($[D_{sr_i}, D_{r_id}] = [0.5, 0.5]$) and the source can select these partners without any limitation. The dependency of the problem to $P_T$, $N_0$ and $d_{sd}$ has the form of $\frac{P_T}{N_0 d_{sd}^\alpha}$ and based on this, we combine the impact of these 3 parameters on one parameter by name of total normalized SNR by the form of $\frac{P_T}{N_0 d_{sd}^\alpha}$.

In this simulation, the source selects all possible states and chooses the best state for best performance. Figure 4 shows the optimal number of partners. This figure shows that when $R$ is increased, the optimal number of partners is decreased and when the total normalized SNR is increased, this number is increased too.

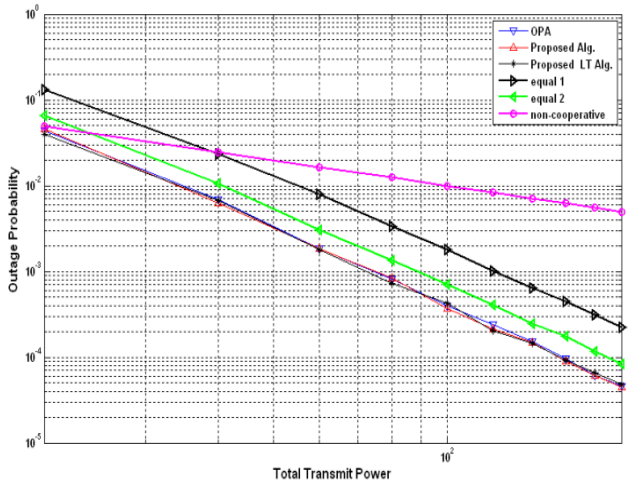

Figure 2- The outage probability of the difference schemes for different total power with two partners

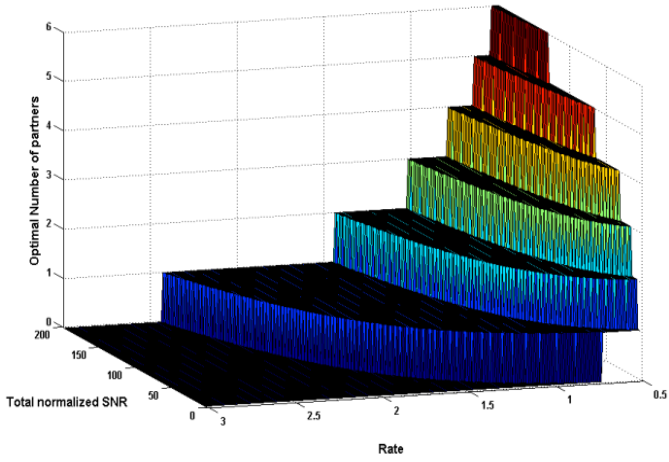

Figure 4- The optimal number of cooperative partners

The above figure shows that when $R$ is greater than a constant rate, by this range of the total normalized SNR, the non-cooperative scheme is preferred to cooperative schemes.

If the locations of the partners have been changed, this figure has been changed with reserving the expressed behavior.

## VI. PARTNER SELECTION

As shown in figure 4, transmission with one relay is preferred to other schemes (greater than one partner and non-cooperative scheme) in great range of $R$ and total normalized SNR. In addition, one of the most important reason to select one relay scheme is MAC and scheduling problems in greater than one relay schemes. For example if the nodes obey from sleep and wakeup discipline, they should sync own sleep program with their partners for cooperation and this is a source of collision if the number of cooperative nodes is increased. For these two reasons, we present the two cooperative partners scheme (one relay for each transmission) in this section.

In two cooperative partners scheme with greedy manner, nodes want to know that with cooperation by one of the other nodes, it can maximize its own performance. In this paper (based on section 3), we assume that the node wants to minimize the outage probability of its transmission by using one cooperative node. For this reason, we must find the outage probability of transmission of one node with one relay in different locations. The approximation of the outage probability of the transmission of one node with one relay is shown in (13). If we extract the effect of network parameters ($P_T$, $R$, $d_{sd}$ and $N_0$) from the outage probability, the term $r_i$ in (14) is obtained.

$$P_{out-2s} \cong \frac{(2^{2R}-1)^2}{2!} * \frac{\left(1+\frac{\lambda_i' D_{r_id}^\alpha}{1-\lambda_i' D_{sr_i}^\alpha}\right)^2}{\left(\frac{P_T}{N_0 d_{sd}^\alpha}\right)^2 \lambda_i'} \quad (13)$$

$$r_i = \left(1+\frac{\lambda_i' D_{r_id}^\alpha}{1-\lambda_i' D_{sr_i}^\alpha}\right)^2 * \frac{1}{\lambda_i'} \quad (14)$$

The node can rank the candidate nodes for cooperation with $r_i$ because the extracted term is equal for all nodes in different locations. This means that if $r_i$ of the i[th] user is smaller than $r_j$ of the j[th] user, the i[th] user is preferred to the j[th] user for cooperation. The amount of $r$ for different locations is shown in figure 5. In this figure, source and destination are placed in $(x,y)=(0,0)$ and $(x,y)=(1,0)$ respectively. The users can select the partners for relaying based on this figure.

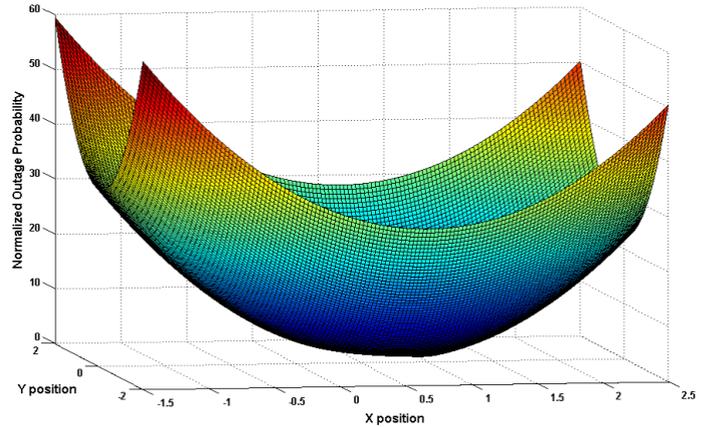

Figure 5- the parameter $r$ for ranking the candidate partner for relaying

If the node wants to calculate the approximated value of the outage probability based on this figure, it should find the value of the location of the cooperative node in figure 5 and multiply it by extracted term of (13).

This method present a greedy manner in network because in selection of partners, all nodes assume the improvement of own transmission and neglect the improvement of performance of the network.

## VII. CONCLUSION

In this paper we presented two power allocation algorithms for Amplify-and-Forward cooperative diversity with a constraint on total transmit power. The proposed algorithms do not require iterative operations to reach the results and can be easily implemented. We demonstrated that the performance of the proposed algorithm closely matches that of the optimal, more complex iterative solutions previously reported in the literature. We evaluated the issue of partner selection in AFCD and showed that the optimal number of cooperative partner is a function of the network parameters and the state of the candidate partners set. We also presented a method for ranking the candidate nodes for cooperation in case on one partner selection which by this method the nodes can select the best partner between all candidate partners for minimizing the outage probability of transmission to destination.

REFERENCES

[1] A. Sendonaris, E. Erkip, and B. Aazhang, "User cooperation diversity–part I: system description," IEEE Trans. Commun., vol. 51, no. 11, pp.1927–1938, Nov. 2003.

[2] J. N. Laneman, D. N. C. Tse, and G. W. Wornell, "Cooperative diversity in wireless networks: efficient protocols and outage behavior," IEEE Trans Inform. Theory, vol. 50, no. 12, pp. 3062-3080, Dec. 2004.

[3] J. N. Laneman and G. W. Wornell, "Distributed space–time coded protocols for exploiting cooperative diversity in wireless networks," IEEE Trans. Inf. Theory, vol. 49, no. 10, pp. 2415–2525, Oct. 2003.

[4] J. Luo, R. S. Blum., L. J. Cimini, L. J. Greenstein and A. M. Haimovich, "Decode-and-Forward Cooperative Diversity with Power Allocation inWireless Networks," in Proc. IEEE Globecom 2005, St Louis, MO, USA, Nov.-Dec. 2005.

[5] R. Annavajjala, P. C. Cosman, L. B. Milstein, "Statistical channel knowledge-based optimum power allocation for

relaying protocols in the high SNR regime," IEEE Journal on Selected Areas in Communications, vol. 25, no. 2, pp.292–305, Feb. 2007.

[6] Y. Zhao, R. Adve, and T. Lim, "Improving amplify-and-forward relay networks: optimal power allocation versus selection," in Proceedings on the IEEE International Symposium on Information Theory (ISIT), 2006.

[7] A. Bletsas, A. Khisti, D.P. Reed, A. Lippman, "A Simple Cooperative. Diversity Method based on Network Path Selection", IEEE Journal on Selected Areas in Communications, vol. 24, no. 3, pp.659–672, Mar. 2006.

[8] A. Nosratinia, T. E. Hunter, "Grouping and partner selection in cooperative wireless networks," IEEE Journal on Selected Areas in Communications, vol. 25, no. 2, pp. 369-378, Feb. 2007.

[9] N. Ahmed, M.A. Khojastepour, B. Aazhang, "Outage Minimization and Optimal Power Control for the Fading Relay Channel," IEEE Information Theory Workshop, San Antonio, TX, October 24-29, 2004.

[10] C. Fischione, A. Bonivento, A. Sangiovanni-Vincentelli and K. H. Johansson, "Cooperative Diversity with Disconnection Constraints and Sleep Discipline for Power Control in Wireless Sensor Networks., in Proc. of IEEE 64th Semiannual Vehicular Technology Conference - Spring 2006 (IEEE VTC Spring 06), Melbourne, Australia, May 2006.

[11] T. Himsoon, W.P. Siriwongpairat, Z. Han, and K.J.R. Liu, "Lifetime Maximization via Cooperative Nodes and Relay Deployment in Wireless Networks", *IEEE Journal of Selected Areas in Communications*, Special Issue on Cooperative Communications and Networking, vol 25, no 2, pp.306-317, Feb 2007.

[12] V. Mahinthan, L. Cai, J. W. Mark and X. Shen, "Maximizing Cooperative Diversity Energy Gain for Wireless Networks," IEEE Trans. Wireless Commun., vol. 6, no. 7, pp.2530–2539, Jul. 2007.

[13] E. Yazdian, M.R. Pakravan, "Adaptive Modulation Technique for Cooperative Diversity in Wireless Fading Channel," PIMRC, Sep 2006.

[14] R. Fletcher. "Practical Methods of Optimization," John Wiley & sons.

[15] J. Nocedal and S. J. Wright, "Numerical Optimization," Springer press.

[16] H. Goudarzi and M.R. Pakravan, "Optimal partner selection and power allocation for amplify and forward cooperative diversity," in proceeding of PIMRC 2008.

[17] H. Goudarzi and M.R. Pakravan, "Equal Power Allocation Scheme for Cooperative Diversity," in proceeding of IEEE/ IFIP international conference on Internet, ICI 2008.

[18] J. N. Laneman, "Cooperative Diversity in Wireless Networks: Algorithms and Architectures," PHD thesis dissertation, Aug. 2002.